\documentclass[useAMS]{mn2e}
\usepackage{graphicx}

\title[Dust formation by WR\,19 at periastron passage]
{Dust formation by the colliding-wind WC5+O9 binary WR\,19 at periastron passage\thanks{Based 
on observations collected at the European Southern Observatory, La Silla, Chile, 
in allocations 67.D-0043, 70.D-0266, 72.D-0070, 73.D-0027, 74.D-0116, 76.D-0029, 
78.D-0001, 79.D-0092 and 80.D-0198.}}
\author[P. M. Williams et al.]
       {P. M. Williams$^1$\thanks{Email: pmw@roe.ac.uk}, 
        G. Rauw$^2$ and K. A. van der Hucht$^{3,4}$\\
   $^1$Institute for Astronomy, Scottish Universities Physics Alliance, 
     University of Edinburgh, Royal Observatory, Edinburgh EH9 3HJ\\
   $^2$Institut d'Astrophysique et de G\'eophysique, Universit\'e de Li\`ege, 
      All\'ee du 6 Ao\^ut 17, B\^at. B5c, 4000 Li\`ege, Belgium\\
   $^3$Space Research Organization Netherlands, Sorbonnelaan 2,
          NL-3584\,CA Utrecht, The Netherlands\\
   $^4$Astronomical Institute `Anton Pannekoek', University of
          Amsterdam, Kruislaan 403, NL-1098 SJ Amsterdam, The Netherlands \\
   }

\date{Accepted 2009 Feb 23.
      Received 2009 Feb 23;
      in original form 2009 Jan 30}

\pagerange{\pageref{firstpage}--\pageref{lastpage}}
\pubyear{2009}

\begin{document}

\maketitle

\label{firstpage}

\begin{abstract}
We present infrared photometry of the episodic dust-making Wolf-Rayet system 
WR\,19 (LS\,3), tracking its fading from a third observed dust-formation 
episode in 2007 and strengthening the view that these episodes are periodic 
($P = 10.1\pm0.1 y$). 
Radial velocities of the O9 component observed between 2001 and 2008 
show RV variations consistent with WC\,19 being a spectroscopic binary of 
high eccentricity ($e=0.8$), having periastron passage in 2007.14, shortly
before the phase of dust formation. In this respect, WR\,19 resembles the 
archetypical episodic dust-making colliding-wind binary system WR\,140.
\end{abstract}


\newpage

\section{Introduction}

\noindent The dense, supersonic winds that give WR stars their 
characteristic emission-line spectra carry significant mass loss 
($\sim 10^{-5}$ M$_{\odot }\,$y$^{-1}$) and kinetic energy 
($\sim 10^{4}$ L$_{\odot }$). The release of some of this energy from 
the collision of such a wind with that of a massive companion in a 
colliding-wind binary (CWB) system gives rise to a range of theoretically 
predicted (X-ray emission) and unpredicted (non-thermal radio emission 
and dust formation) phenomena. 
The association of dust formation with colliding winds began with the 
demonstration that the 2900-d periodic dust formation episodes by the 
archetypal Wolf-Rayet colliding-wind binary WR\,140 occurred during 
periastron passages of its highly eccentric orbit (Williams et al. 1990a).
The high densities (10$^{3}$-- 10$^{4}$ times that of the undisturbed 
Wolf-Rayet wind) required for dust formation to occur can be produced in 
colliding-wind shocks if they cool efficiently (Usov 1991). The link 
between the dust-formation episodes and binary orbit in WR\,140 is 
provided by periodic increases of the {\em pre-shock} wind density by a 
factor of $\sim$ 40 for a brief time during periastron passage when the 
separation of the WC7 and O5 stars is at a minimum (Williams 1999). 
Slightly different behaviour is shown by the WC7+O9 periodic dust-maker 
WR\,137, whose dust-formation and RV orbital periods are identical within 
the uncertainties, but there is a 1.3-y (0.1~P) delay between periastron 
passage and infrared maximum (Williams et al. 2001, Lef\`evre et al. 2005). 
Evidence for a CWB origin for the persistent dust formation by many 
WC8--9 stars comes from the rotating `pinwheel nebulae' observed around 
WR\,104 (Tuthill, Monnier \& Danchi 1999) and WR\,98a (Monnier, 
Tuthill \& Danchi 1999) -- although it should be noted that we do not 
have orbits for these systems, and only WR\,104 has a spectroscopic 
companion. 

These results show the way to solving the long-standing mystery of dust 
formation by Wolf-Rayet stars within the framework of wind compression and 
cooling in CWBs. The processes are being intensively studied in WR\,140, 
whose orbit is now well defined (Marchenko et al. 2003, Dougherty et al. 2005) 
and whose dust has been imaged at high resolution 
(Monnier, Tuthill \& Danchi 2002, Williams et al. 2007), but further examples 
are needed where we can relate the dust formation to the binary orbit. 

For this purpose, we selected WR\,19 (= LS\,3, Smith 1968), which differs 
from other dust-making WR stars in having an earlier spectral subtype. 
In her discovery paper, Smith classified its spectrum as WC5+OB, the `+OB' 
inferred from the weakness of the emission lines (footnote in Smith, 
Shara \& Moffat 1990a, who noted the absence of absorption lines). 
It was reclassified as a member of the new WC4 sub-class in the Sixth 
Catalogue (van der Hucht et al. 1981) but was returned to WC5 by Crowther, 
De Marco \& Barlow (1998), in both cases without reference to a companion. 
In either event, the subtype is earlier than those of the other episodic and 
persistent dust makers (WC7--8 and WC8--10 respectively).
Dust formation by WR\,19 was first reported by Williams et al. (1990b, 
hereafter Paper~1), who found a near-infrared spectral energy distribution 
(SED) showing 780-K dust emission, which evolved to one characteristic of 
the stellar wind within two years as the dust emission faded. 
This prompted continued infrared monitoring to look for another episode of 
dust formation, and spectroscopy to search for the companion suggested by 
the weak emission lines (Smith) and possible CWB origin of the dust. 
The results of both searches were reported by Veen et al. (1998, hereafter 
Paper~2), who discovered a second dust-formation episode 10.1~y after the 
first and presented blue-green spectra showing absorption lines from which 
the companion was classified as a O9.5--9.7 star. They concluded that 
WR\,19 was probably an eccentric WCE+O9.5--9.7 binary. 

If the WR\,140 paradigm held for WR\,19, we expected it to be a spectroscopic 
binary of fairly high orbital eccentricity having its next periastron passage 
coinciding with its next dust formation episode in 2007--08. We 
therefore set out to observe its RV to look for variations to confirm its 
status as a binary, continuing at least until after 2008. We also sought 
infrared photometry to confirm the expected dust-formation episode and 
apparent 10.1-y period inferred from the first two episodes observed.
In this paper, we report these observations and the confirmation of the 
CWB status of WR\,19.

\section{Observations}\label{SObs}      

\begin{figure}                                              
\includegraphics[clip,width=8cm]{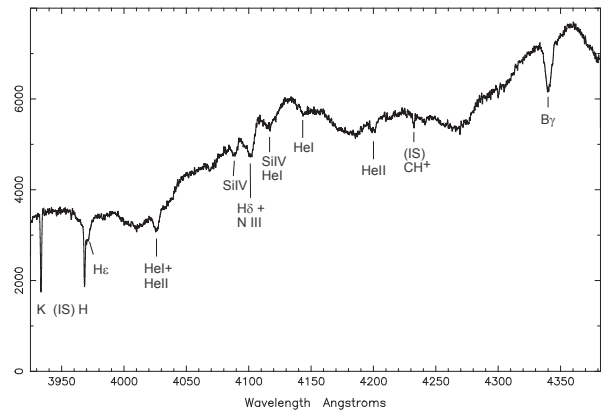}
\caption{Mean spectrum of WR\,19 compiled from spectra observed between 2003 and 2005.}
\label{Fspec}
\end{figure}

The spectra were observed with the EMMI instrument on the 3.5-m New Technology 
Telescope (NTT) at the European Southern Observatory, La Silla. As the 
investigation required relatively short observations spread over several years, 
all but the first were taken in the Service Observing mode, and we continued 
seeking observations for as long as this programme was offered at La Silla. 
Fortunately, this continued long enough to take us through periastron passage. 
We elected to concentrate our search for RV variations on the absorption-line 
spectrum. We used the EMMI BLMD Grating 
\#3 and set it to give a spectrum running  from 3925\AA\ to 4382\AA\, 
(Fig.\,\ref{Fspec}) covering the interstellar Ca\,{\sc ii} K line to H$\gamma$. 
The interstellar lines were included to provide a check on the wavelength scale. 
An 0.7-arcsec entrance slit gave a spectral resolution of 1.06~\AA\ (2.5 pixels), 
confirmed from measurement of the interstellar K line. 
A standard Observing Block (OB) comprising two 1475-s integrations on the star 
followed by 240~s on the ThAr calibration lamp (separate Th and Ar lamps for the 
earliest observations) was compiled to fit the 1-h OB limit for service observations. 
Inevitably, there was a large range in S/N ratio (typically 40--80) in the 
final spectra depending on the observing conditions, and sometimes two OBs were 
observed sequentially if conditions were particularly poor. 

\begin{figure}                                    
\includegraphics[clip,width=8cm]{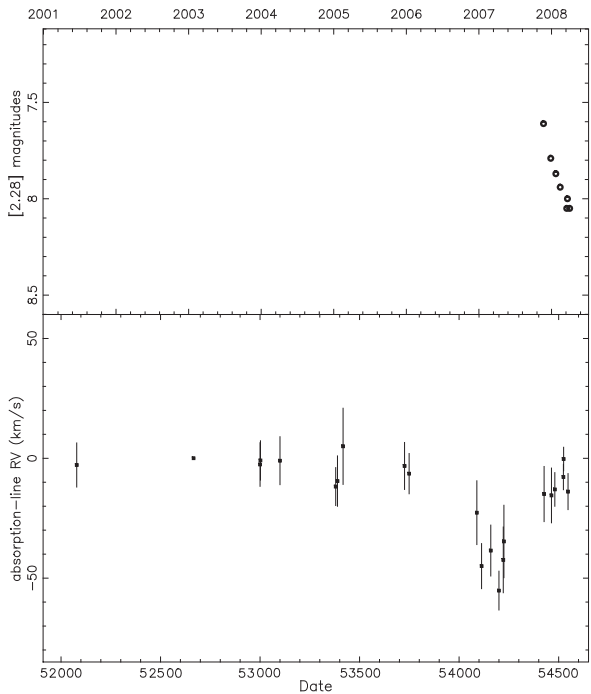}
\caption{Infrared [2.28] magnitudes (top) and radial velocities (bottom) of 
WR\,19 plotted against date (civil and MJD). The velocities are relative to that 
in 2003 January. The IR flux is observed to be fading as the system moves away 
from periastron. The pre-outburst flux level is expected to be [2.28]$\simeq$8.84.}
\label{SofIRV}
\end{figure}

The spectra were reduced using {\sc figaro}. Initial experiments suggested that 
the wavelength shifts of the emission features available (Fig.\,\ref{Fspec}) 
could be determined by cross-correlation with a template derived from a high-quality 
spectrum (that observed in 2003 January), but this was found not to be the case 
when more data became available. Instead, we used the cross-correlation method to 
derive the absorption-line velocities, useful in dealing with blends like that 
of He\,{\sc ii} with He\,{\sc i} at 4026\AA\ and H$\delta$ with N\,{\sc iii} at 
4097\AA\ and 4103\AA, which broadened the profile to 7.3\AA\ compared with 5.7\AA\ 
for H$\gamma$. 
The use of a template based on a spectrum of the object star and rectification 
of the spectral regions covering each of the absorption lines before cross-correlation 
minimised the introduction of systematic errors in derivation of the RVs.
We measured velocities for the absorption lines individually, from which we derived 
uncertainties, and separated those of the interstellar Ca\,{\sc ii} H line from 
H$\epsilon$ by constructing two versions of the template and appropriate masking. 
Finally, small adjustments to the velocities were made to fit the shifts of the 
interstellar Ca\,{\sc ii} H and K lines and, where possible, CH$^+$ at 4232\AA. 
The relative RVs and their errors are given in Table \ref{TRV} and plotted against 
date in Fig.\,\ref{SofIRV}. 
We also measured the heliocentric RV of the 2003 January spectrum used for 
the template by fitting Gaussians to the absorption lines (apart from the 
H$\delta$+N\,{\sc iii} blend) to be  8.6$\pm$5.4 km~s$^{-1}$.

\begin{table}
\caption{Absorption-line radial velocities (km~s$^{-1}$) of WR\,19 relative 
to that on MJD 52666 (2003 January 26). 
The phases were calculated using P = 3689~d. from the photometry and $T_{\rmn 0}$ 
= MJD 50500 from the orbital solution.}
\begin{tabular}{lcrrr}
\hline
MJD    & $\phi$ &  RV  &  s.d. & (O-C) \\
\hline
52079  &  0.43  &  -2.8 &  9.9 &  -1.3 \\
52666  &  0.59  &   0.0 &  0.0 &   2.0 \\
53000  &  0.68  &  -2.6 &  9.1 &   0.3 \\
53002  &  0.68  &  -0.9 &  8.2 &   2.0 \\
53100  &  0.70  &  -1.0 & 10.0 &   1.9 \\
53380  &  0.78  & -11.8 &  7.9 &  -7.7 \\
53389  &  0.78  &  -9.5 & 10.5 &  -5.0 \\
53417  &  0.79  &   5.0 & 15.9 &   9.5 \\
53726  &  0.87  &  -3.2 &  9.8 &   4.5 \\
53749  &  0.88  &  -6.4 &  8.4 &   1.6 \\
54089  &  0.97  & -22.7 & 13.3 &   5.2 \\
54113  &  0.98  & -45.0 &  9.4 & -10.8 \\
54159  &  0.99  & -38.5 & 10.6 &   8.9 \\
54200  &  1.00  & -55.2 &  8.1 &  -4.5 \\
54222  &  1.01  & -42.4 & 13.7 &   1.4 \\
54225  &  1.01  & -34.7 & 15.1 &   7.7 \\
54427  &  1.06  & -14.9 & 11.5 &  -4.2 \\
54464  &  1.07  & -15.5 & 11.4 &  -6.1 \\
54481  &  1.08  & -13.0 &  7.0 &  -4.2 \\
54524  &  1.09  &  -7.8 &  5.3 &  -0.3 \\
54525  &  1.09  &  -0.3 &  4.9 &   7.2 \\
54547  &  1.10  & -13.9 &  7.5 &  -6.8 \\
\hline
\end{tabular}
\label{TRV}
\end{table}

The near-IR photometry (Table \ref{TPhot}) was derived from images observed with 
the SofI instrument on the NTT in the large-field (4.9 arcmin) configuration. 
To avoid saturating the detector, we observed WR\,19 through the narrow-band 
($\Delta\lambda = 0.03\umu$m) 2.28-$\umu$m filter. Each observation comprised an 
11-point autojitter each of 3$\times$1.5-s snapshots. The median of these images 
was used to flat-field the frames. Four stars in the field with high-quality 
$2MASS$ (Skrutskie et al. 1990) data bracketing WR\,19 in $K_{\rmn{s}}$ and 
having a range of $(H-K_{\rmn{s}})$ were used to calibrate the images. 
The central wavelength of the $K_{\rmn{s}}$ filter (2.16$\umu$m) is shorter than 
that of the [2.28] filter, and we used the observations of the calibrators to 
derived a colour equation:
\[
[2.28] = K_{\rmn{s}} - 0.25 (H-K_{\rmn{s}}).
\]
From this, we derived the [2.28] magnitudes of the calibrators, and then of WR\,19. 
The latter are given in Table \ref{TPhot} and plotted in Fig.\,\ref{SofIRV}. 
The colour equation does not apply to the magnitudes of WR\,19 itself owing to the 
significant contribution to the $K$- and $K_{\rmn{s}}$-band fluxes of WC stars 
from the strong emission lines in their spectra (e.g. Williams 1982) and we consider 
this below.

\begin{table}
\caption{Narrow-band [2.28] magnitudes of WR\,19. The phases were calculated using 
P = 3689~d. and $T_{\rmn 0}$ = MJD 50500.}
\centering
\begin{tabular}{lcc}
\hline
MJD    &$\phi$ & [2.28] \\
\hline
54427  & 1.06  &  7.61  \\
54463  & 1.07  &  7.81  \\
54464  & 1.07  &  7.76  \\
54487  & 1.08  &  7.87  \\
54511  & 1.09  &  7.94  \\
54544  & 1.10  &  8.05  \\
54547  & 1.10  &  8.00  \\
54557  & 1.10  &  8.05  \\
\hline
\end{tabular}
\label{TPhot}
\end{table}  

\begin{figure}                                         
\includegraphics[clip,width=8cm]{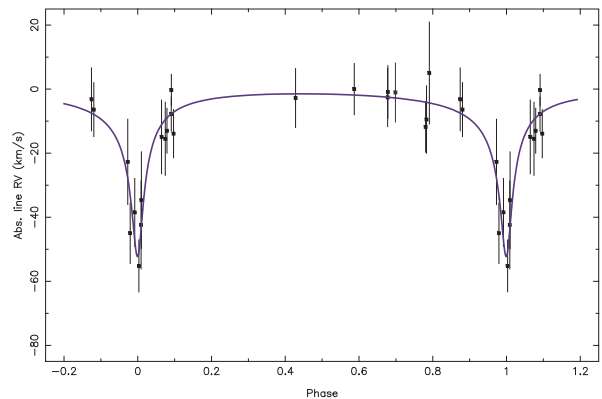}
\caption{Absorption-line radial velocities and orbit solution plotted against 
phase.}
\label{RVorbit}
\end{figure}

\section{Results} 
\label{SResults} 

The RVs and [2.28] mags are plotted on the same time-scale in Fig.\,\ref{SofIRV}. 
The velocities show little variation until 2007, when they go through a minimum 
characteristic of periastron passage in a binary orbit. 
While the velocities were recovering, the photometry 
shows rapid fading. We do not have photometry during the RV sequence prior to 
November 2007, but the photometry in Papers 1 and 2 show an average $K$ = 8.55, 
much fainter than any of our [2.28] mags, so these results confirm our expectation 
that WR\,19 is a CWB with dust-formation coinciding with periastron passage.

The [2.28] magnitudes are a good measure of the stellar continuum but $K$ magnitudes 
observed at the same time would have been brighter owing to the contribution from 
strong emission lines in the broader $K$ filter passband. From low-resolution 
infrared spectroscopy of WR\,19, Smith \& Hummer (1988) derived an emission-line 
correction of $\Delta K = 0.29$ mag. Their spectrum was observed in 1982 May, 
when the dust emission would have been negligible according to our light curve, 
so the dust-free WR\,19 should have [2.28] $\simeq$ 8.84. Also, we can use 
Smith \& Hummer's measured equivalent widths (totalling 0.13$\umu$m in 
the $K$ passband) to estimate the $K$ magnitudes corresponding to our [2.28] 
observations, assuming the emission-line fluxes to be constant. 
The emission-line corrections in 2007--08 are smaller than that in 1982 
determined by Smith \& Hummer owing to the additional contribution of the dust 
emission at this time, and range from 0.10 to 0.15 mag. as [2.28] faded from 
7.61 to 8.05, giving `corrected' $K$ fading from 7.51 to 7.90 in 2007--08. 
Unfortunately, there is no overlap between these and the $K$ magnitudes 
observed after the 1988 and 1998 episodes, so we cannot use the new observations 
to refine the period, but phased light curves using different trial periods and 
extrapolating the 2007--08 fading showed the new photometry to be consistent 
with a period equal to the interval of 10.1$\pm$0.1~y between the 1988 and 1998 
events (Paper~2). This confirms the periodicity of the dust-formation episodes, 
and observations of the dust-formation episode expected in 2017 would be valuable 
to reduce the 0.1-y uncertainty.

\begin{table}
\centering
\caption{Orbital elements for the O9 component of WR\,19 from 
RVs with period fixed to the photometric period.}
\begin{tabular}{cl}
\hline
Parameter     &  Value           \\
\hline
P             &  3689 d. (fixed) \\		
T$_{\rmn{0}}$ & MJD 50500$\pm$14 \\
e             &  0.80$\pm$0.04  \\
$\omega$      &  184$\degr\pm$4$\degr$ \\
K             &  25$\pm$2 km~s$^{-1}$ \\
$\gamma$      &  2$\pm$5 km~s$^{-1}$ \\
a$\sin(i)$    &  5.3$\pm$0.6 au  \\
f(m)          &  1.3$\pm$0.8 M$_{\odot}$ \\
\hline
\end{tabular}
\label{TElIS}
\end{table}

We then solved the radial velocity orbit with the period fixed to the photometric 
period, and derived the elements in Table \ref{TElIS}. 
A nominal error of 8 km~s$^{-1}$ was attached to the template velocity to avoid 
over-weighting this point in the RV solution. The RV curve is plotted 
over the individual velocities in Fig.\,\ref{RVorbit} and the differences from 
the orbit (O-C) tabulated with the observations in Table \ref{TRV}. Certainly, 
it would be desirable to have a longer run of RV measurements covering two 
periastron passages to measure an RV period and test its agreement with that of 
the photometry, but the present result indicates a high eccentricity orbit with 
dust-formation at periastron passage. 
The $\gamma$-velocity in Table \ref{TElIS} has been converted to a heliocentric 
velocity using that of the template (MJD 52666) spectrum.

\begin{figure}                                      
\includegraphics[clip,width=8cm]{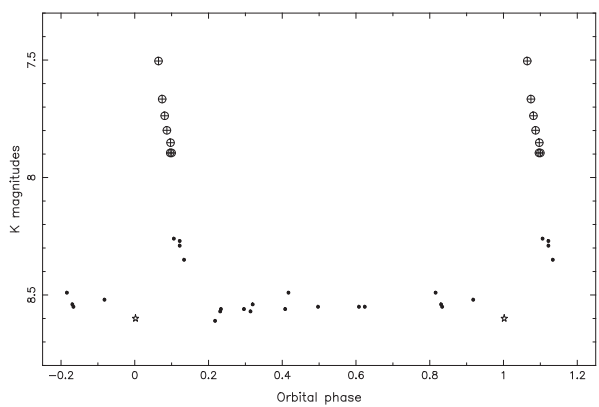}
\caption{Light curve of WR\,19 with $K$ magnitudes from Papers 1 and 2 ($\bullet$), 
converted from [2.28] ($\oplus$), and from the DENIS survey ($\star$).}
\label{PhasedK}
\end{figure}

The $K$-band photometry, including $K$ converted from [2.28], is plotted against 
orbital phase in Fig.\,\ref{PhasedK}. The converted $K$ magnitudes in 2007--08 
show WR\,19 to be brighter in the near-IR than any of the previous $K$ photometry 
(Papers 1 and 2), indicating that we observed WR\,19 closer to its infrared maximum 
than before, but we still need to observe the maximum and the rise to it, next 
due in 2017. The rate at which $K$ faded, 1.2~mag~y$^{-1}$, is slightly less that 
of WR\,140 in its early stages, 1.4~mag~y$^{-1}$, and suggests a slightly slower 
movement of dust away from the stars heating it. The terminal wind velocity of 
WR\,19 needs to be measured to examine this further; radiation pressure from the 
stars also plays a role by accelerating the dust relative to the wind until it 
reaches a constant drift velocity and this effect may be smaller in WR\,19 than 
that in WR\,140 owing to the later spectral type of the O companion.

Fortuitously, the field including WR\,19 in the DENIS survey (Epchtein et al. 1999) 
was observed almost exactly at the time of the 1997 periastron passage 
(MJD 50506, $\phi$ = 0.996). This observation ($K$ = 8.58, close to the mean 
dust-free level of $K$ = 8.55) shows no evidence for dust emission. 
On the other hand, the first [2.28] observation shows that dust emission was already 
fading at phase 0.06, i.e. that the dust was cooling as the stellar wind carried it 
away from the stars heating it, and it was no longer being replenished by nucleation 
of fresh dust. This constrains the interval during which dust was nucleating to 0.06~P 
(220~d) at most, with a similar upper limit on the delay between periastron passage 
and $K$-flux maximum. This time-scale is comparable to, or slightly slower than, that 
of WR\,140 (0.03~P delay, Williams 1999), consistent with its slightly lower orbital 
eccentricity ($e = 0.8$, compared with $e=0.88$ for WR\,140, Marchenko et al. 2003), 
and the idea that dust formation is triggered when the separation of the binary 
components in their orbit falls below a critical value, leading to a change in the 
wind-collision process.

We also combined the spectra observed in 2003--05 to form a higher S/N spectrum 
(Fig.\,\ref{Fspec}) than that used in Paper~2 to re-examine the spectral type of the 
companion. The wavelength range does not include the `formal' O-star classification 
lines, but we compared the relative strengths of the $\lambda$ 4200\AA\ He\,{\sc ii} 
line and visible He\,{\sc i} lines with those in Walborn \& Fitzpatrick's (1990) 
atlas and consider the spectral type of the companion to be earlier than O9.5 and 
closer to O9, which we adopt.

The derived mass function allows us to estimate a minimum mass for the WC star, 
11$\pm$2 M$_{\odot }$ assuming 20 M$_{\odot }$ for the O9 star. This is consistent 
with the range of masses of WC stars found from binary orbits (9--16 M$_{\odot }$, 
van der Hucht 2001). It also suggests that the orbit of WR\,19 may be fairly highly 
inclined unless the mass of the WC star is unusually high.

\section{Discussion} 
\label{SDisc} 

\begin{table}
\centering
\caption{Properties of WR\,19. including photometry of the dust-free wind.}
\begin{tabular}{lll}
\hline
Parameter                &  Value        &  Reference  \\
\hline
Spectrum                 &  WC5 + O9     & Crowther et al. + this work  \\
Distance                 &  1.7--3.9 kpc & Smith, Shara \& Moffat (1990b) \\		
A$_v$ (1.1A$_{\rmn{V}}$) & 5.6           & Smith et al. (1990b) \\
$v$                      & 13.85         & Smith (1968) \\
$J$                      & 9.78          & Papers 1 \& 2 \\
$K$                      & 8.55          & Papers 1 \& 2 \\
$L^{\prime}$             & 8.20          & Papers 1 \& 2 \\
$[8.0]$                  & 7.20          & $GLIMPSE$  \\
\hline
\end{tabular}
\label{TProps}
\end{table}  

The new observations of WR\,19 have confirmed its status as a periodic, dust-forming 
CWB. More observations are needed to strengthen the period and orbit, including the 
WC star to get a mass ratio, but WR\,19 is available as a laboratory to study wind 
collision effects. For example, spectroscopy of the He\,{\sc i} $\lambda$1.083-$\umu$m 
line will be valuable not only to determine the wind velocity but also for mapping the 
wind-collision region through variation of the absorption component and the appearance 
and movement of sub-peaks on the broad emission component, as in WR\,140 (Varricatt, 
Williams \& Ashok 2004).

Leitherer, Chapman \& Koribalski (1997) and Chapman et al. (1999) included WR\,19 
in their surveys of radio emission from southern WR stars but found only upper limits. 
This does not rule out non-thermal emission from colliding winds, because the longer 
wavelength (13 and 20~cm) observations, more likely to show non-thermal emission 
having a negative spectral index, were taken in 1997.15, very close to periastron 
passage when the wind-collision region would have been most deeply embedded in the 
stellar winds and the circumstellar free-free extinction greatest -- the non-thermal 
emission from WR\,140 is extinguished at this phase. Re-observation of WR\,19 at 
different phases may reveal non-thermal emission when the geometry is more favourable.
We can estimate the radio flux density of WR\,19's stellar wind by assuming that its  
spectral index between mid-IR and cm wavelengths is similar to those of the 
WC5 stars, WR\,111 and WR\,114, observed at 6~cm and 3.6~cm by Cappa, Goss \& van der 
Hucht (2004) and Bieging, Abbott \& Churchwell (1982) respectively. 
For the mid-IR fluxes of the stellar wind, we use the $GLIMPSE$ (Benjamin et al. 2003) 
observations of WR\,19, taken at phase 0.68 (MJD 52997) when the dust emission is 
assumed to be long since faded; the $GLIMPSE$ [3.6] and [4.5] magnitudes are in 
excellent agreement with our dust-free $L$, $L^{\prime}$ and $M$ magnitudes.
To scale the radio flux densities of WR\,111 and WR\,114, we used the GLIMPSE [8.0] 
magnitudes of the three stars, leading to an estimated flux densiteis of $\sim$ 0.07 mJy 
at 3.6 and 6~cm for WR\,19. These are about one-quarter the 3-$\sigma$ upper limit 
reported by Leitherer et al., so observation of the WR\,19 stellar wind should be 
possible in a reasonable integration time, providing a baseline for studying any 
non-thermal radio emission. 

Similarly, although WR\,19 was not detected in the {\em ROSAT} survey of WR stars 
(Pollock, Haberl \& Corcoran 1993), its confirmation as a CWB and knowledge of its 
orbit justify re-observation of WR\,19 at X-ray wavelengths. Selected properties of 
WR\,19 are collected in Table \ref{TProps}.

\section*{Acknowledgments}

All observations apart from the first spectrum were taken in the Service 
Observing mode, and it is a pleasure to thank the NTT support astronomers 
at La Silla Observatory for their care over the years in taking the observations 
and helpful correspondence regarding their preparation.
This publication makes use of data products from the Two Micron All Sky 
Survey, which is a joint project of the University of Massachusetts and 
the Infrared Processing and Analysis Center/California Institute of 
Technology, funded by the National Aeronautics and Space Administration 
and the National Science Foundation.

\end{document}